\documentclass[aps,prd,showkeys,showpacs,twocolumn,superscriptaddress]{revtex4}

\usepackage[dvips]{graphics}

\usepackage{amsfonts}
\usepackage{amsmath}
\usepackage{latexsym}
\usepackage{amssymb}
\usepackage[mathcal]{euscript}

\newtheorem{Post}{Postulate}
\newtheorem{Pro}{Property}
\newtheorem{Pb}{Problem}
\newtheorem{Algo}{Algorithm}

\begin{document}

\title{\begin{center}
Quantum Computation explained to my Mother
\end{center}
}

\author{Pablo Arrighi}
\email{pja35@cam.ac.uk} \affiliation{Computer Laboratory,
University of Cambridge, 15 JJ Thomson Avenue, Cambridge CB3 0FD,
U.K. }



\begin{abstract}
There are many falsely intuitive introductions to quantum theory
and quantum computation in a handwave. There are also numerous
documents which teach those subjects in a mathematically sound
manner. To my knowledge this paper is the shortest of the latter
category. The aim is to deliver a short yet rigorous and
self-contained introduction to Quantum Computation, whilst
assuming the reader has no prior knowledge of anything but the
fundamental operations on real numbers. Successively I introduce
complex matrices; the postulates of quantum theory and the
simplest quantum algorithm. The document originates from a fifty
minutes talk addressed to a non-specialist audience, in which I
sought to take the shortest mathematical path that proves a
quantum algorithm right.
\end{abstract}

\maketitle
\section{Some mathematics}
I will begin this introduction with less than three pages of
mathematics, mainly definitions. These notions constitute the
vocabulary, the very language of quantum theory, and every single
one of them will find its use in the second part, when I introduce
the postulates of quantum theory.

\subsection{Complex Numbers}
A \emph{real} number is a number just like you are used to. E.g.
$1,\; 0,\;-4.3\;$ are all real numbers. A \emph{complex} number,
on the other hand, is just a pair of real numbers. I.e. suppose
$z$ is a complex number ($z$ is just a name we give to the number,
we could call it \emph{zorro}), then $z$ must be of the form $(a,b)$ where $a$ and $b$ are real numbers.\\
Now I must teach you how to add or multiply complex numbers.
Suppose we have two complex numbers $z_1=(a_1,b_1)$ and
$z_2=(a_2,b_2)$. Addition first: $z_1+z_2$ is defined to be the
pair of real numbers $(a_1+a_2,b_1+b_2)$. And now multiplication
(when I put two number next to one another, with no sign in
between that means they are multiplied): $z_1 z_2$ is defined to
be the pair of real numbers given
by $(a_1 a_2-b_1 b_2,a_1 b_2+a_2 b_1)$. \\
Sometimes we want to change the sign of the second (real)
component of the complex number $z$. This operation is called
\emph{conjugation}, and is denoted by a upper index `$^*$', i.e.
$z^*$ is defined to be the pair of real numbers $(a,-b)$. \\
Another useful operation we do on a complex number is to take its
\emph{norm}. The norm of $z=(a,b)$ is defined to be the real
number $\sqrt{a^2+b^2}$. This operation is denoted by two vertical
bars surrounding the complex number, in other words $|z|$ is
simply a notation for $\sqrt{a^2+b^2}$.
\subsection{Matrices}
A \emph{matrix} of \emph{things} is a table containing those
things, for instance: $ \left(\begin{array}{cc}
\heartsuit     &\spadesuit \\
\diamondsuit     &\clubsuit
\end{array}\right)
$ is a matrix of card suits.\\
We shall call this matrix $M$ for use in later examples.\\
A matrix does not have to be square. We say that a matrix is
$m\times n$ if it has $m$ horizontal lines and $n$ vertical
lines.\\
For instance a \emph{column} is a $1\times n$ matrix e.g:
$\left(\begin{array}{c}
\heartsuit     \\
\diamondsuit
\end{array}\right) $. \\
Similarly a \emph{row} is a $m\times 1$ matrix, e.g. $
\left(\begin{array}{cc}
\heartsuit     &\spadesuit \\
\end{array}\right)
$ is a row. \\
The $ij$-\emph{component} of a matrix designates the `thing' which
is sitting at vertical position $i$ and horizontal position $j$ in
the table, starting from the upper left corner. For instance the
$2\;1$-component of $M$ is $\diamondsuit$. If $A$ is a matrix then
the $ij$-component of $A$ is denoted $A_{ij}$, e.g. here you have
that $M_{11}=\heartsuit,\;M_{21}=\diamondsuit$ etc. \\
Given a matrix we often need to make vertical lines into
horizontal lines and vice-versa. This operation is called
\emph{transposition} and is written `$^t$'. We thus have
$A^t_{ij}=A_{ji}$, in other words if $A$ the $m\times n$ matrix
with $ij$-component $A_{ij}$, then $A^t$ is defined to be the
$n\times m$ matrix which has $ij$-component $A_{ji}$.  Here are
two examples:
\begin{align*}
M^t=\left(\begin{array}{cc}
\heartsuit     &\diamondsuit \\
\spadesuit     &\clubsuit
\end{array}\right)
\quad;\quad \left(\begin{array}{c}
\heartsuit     \\
\diamondsuit
\end{array}\right)^t=
\left(\begin{array}{cc}
\heartsuit     &\diamondsuit \\
\end{array}\right)
\end{align*}
\subsection{Matrices of Numbers}
Let us now consider matrices of numbers. The good thing about
numbers (real or complex, it does not matter at this point) is
that you know how to add and multiply them. This particularity
will now enable us to define addition and multiplication \emph{of
matrices of these numbers}.\\
In order to add two matrices $A$ and $B$ they must both be
$m\times n$ matrices (they have the same size). Suppose $A$ has
$ij$-components. Then $A+B$ is defined to be the $m\times n$
matrix with $ij$-components $A_{ij}+B_{ij}$.\\
If we now want to multiply the matrix $A$ by the matrix $B$ it has
to be the case that the number of vertical lines of $A$ equals
that of the number of horizontal lines of $B$. Now suppose $A$ is
an $m\times n$ matrix with $ij$-components $A_{ij}$, whilst $B$ is
$n\times r$ and has $pq$-components $B_{pq}$. Then $AB$ is defined
to be the $m\times r$ matrix with $iq$-components $A_{i1}
B_{1q}+A_{i2} B_{2q}+..+A_{in} B_{nq}$.\\
To make things clear let us work this out explicitly for general
$2\times 2$ matrices of numbers:
\begin{align*} \textrm{Let}\quad A= \left(\begin{array}{cc}
A_{11}     &A_{12} \\
A_{21}     &A_{22}
\end{array}\right)\quad\textrm{and}\quad B=\left(\begin{array}{cc}
B_{11}     &B_{12} \\
B_{21}     &B_{22}
\end{array}\right)
\end{align*}
\begin{align*}
\textrm{Then}\quad A+B&= \left(\begin{array}{cc}
A_{11}+B_{11}     &A_{12}+B_{12} \\
A_{21}+B_{21}     &A_{22}+B_{22}
\end{array}\right)\\
\quad \textrm{and}\quad\quad\; AB&= \left(\begin{array}{cc}
A_{11}B_{11}+A_{12}B_{21}     &A_{11}B_{12}+A_{12}B_{22} \\
A_{21}B_{11}+A_{22}B_{21}     &A_{21}B_{12}+A_{22}B_{22}
\end{array}\right)
\end{align*}
\subsection{Matrices of Complex Numbers}
Matrix addition and multiplication work on numbers, whether they
are real or complex. But from now we look at matrices of complex
numbers only, upon which we define one last operation called
\emph{dagger}.\\
To do a dagger operation upon a matrix is to transpose the matrix
and then to conjugate all the complex numbers it contains. This
operation is denoted `$^{\dagger}$'. We thus have
$A^{\dagger}_{ij}=A^*_{ji}$, in other words if $A$ is the $m\times
n$ matrix with $ij$-component $A_{ij}$, then $A^{\dagger}$ is
defined to be the $n\times m$ matrix which has
$ij$-component $A^*_{ji}$.\\
Quite a remarkable $n\times n$ matrix of complex numbers is the
one we call `the \emph{identity} matrix'. It is defined such that
its $ij$-component is the complex number $(0,0)$ when $i\neq j$,
and the complex number $(1,0)$ when $i=j$. The $n\times n$
identity matrix is denoted $I_n$, as in:
\begin{align*}
I_1= \left(\begin{array}{c} (1,0)
\end{array}\right)
\quad \textrm{and}\quad I_2= \left(\begin{array}{cc}
(1,0)     &(0,0) \\
(0,0)     &(1,0)
\end{array}\right)
\end{align*}
Having defined the identity matrices we are now able to explain
what it means to be a \emph{unit} matrix of complex numbers.
Consider $M$ an $m\times n$ matrix of complex numbers. $M$ is said
to be a unit matrix if (and only if) it is true that
$M^{\dagger}M=I_n$.
\subsection{Some properties}
You may skip the following three properties if you wish, but they
will be needed in order to fully understand the comments which
follow postulates \ref{evolution} and \ref{measurement}. Moreover
by going through the proofs you will exercise your understanding
of the many definitions you have just swallowed.

\begin{Pro}\label{identity} Let $A$ be an $n\times m$ matrix of complex numbers
and $I_m$ the $m\times m$ identity matrix. We then have that
$AI_m=A$. In other words multiplying a matrix by the identity
matrix leaves the matrix unchanged.
\end{Pro}
\emph{Proof.} First note that a complex number $(a,b)$ multiplied
by the complex number $(1,0)$ is, by definition of complex number
multiplication, given by $(1a-0b,0a+1b)$, which is just $(a,b)$
again. Likewise note that a complex number $(a,b)$ multiplied by
the complex number $(0,0)$ is given by $(0a-0b,0a+0b)$, which is
just $(0,0)$. Now by definition of matrix multiplication the
$iq$-component of $AI_m$ is given by: (where we denote $I_m$ by
just $I$)
\begin{align*}
(AI)_{iq}&=A_{i1}I_{1q}+A_{i2}I_{2q}+..+A_{in} I_{nq}\\
&=A_{i1}(0,0)+A_{i2}(0,0)+..+A_{iq}(1,0)+..+A_{in}(0,0)
\end{align*}
The second line was obtained by replacing the $I_{pq}$ with their
value, which we know from the definition of the identity matrix.
Now using the two remarks at the beginning of the proof we can
further simplify this equation:
\begin{align*}
(AI)_{iq}&=(0,0)+(0,0)+..+A_{iq}+..+(0,0) \\
&=A_{iq} \quad\textrm{by complex number addition.}
\end{align*}
Thus the components of $AI$ are precisely those of $A$.
$\quad\square$

\begin{Pro}\label{dagger} Let $A$ be an $m\times n$ matrix of
complex numbers and $B$ be an $n \times r$ matrix of complex
numbers. Then the following equality is true:
\begin{align*}
(AB)^{\dagger}=B^{\dagger}A^{\dagger}
\end{align*}
\end{Pro}
\emph{Proof.} First note that
\begin{align}
\label{conj plus}
((a_1,b_1)+(a_2,b_2))^{*}=(a_1,b_1)^{*}+(a_2,b_2)^{*}
\end{align}
This is obvious since
\begin{align*}
((a_1,b_1)+(a_2,b_2))^{*}&=(a_1+a_2,b_1+b_2)^{*} \\
&=(a_1+a_2,-b_1-b_2)\quad\textrm{and}\\
(a_1,b_1)^{*}+(a_2,b_2)^{*}&=(a_1,-b_1)+(a_2,-b_2)\\
&=(a_1+a_2,-b_1-b_2)\quad\textrm{as well.}
\end{align*}
Likewise note that
\begin{align}
\label{conj mult}
((a_1,b_1)(a_2,b_2))^{*}=(a_1,b_1)^{*}(a_2,b_2)^{*}
\end{align}
and also
\begin{align}
\label{commut} (a_1,b_1)(a_2,b_2)=(a_2,b_2)(a_1,b_1)
\end{align}
again this is easily verified by computing the left-hand-side
and the right-hand-side of those equalities. \emph{You may want to check this as an exercise.}\\
Now by definition of matrix multiplication we have that
\begin{align*}
(AB)_{iq}=A_{i1}B_{1q}+A_{i2}B_{2q}+..+A_{in} B_{nq}
\end{align*}
Thus the components of $(AB)^{\dagger}$ are given by
\begin{align*}
(AB)^{\dagger}_{iq}&=(AB)^{*}_{qi}\\
&=A^{*}_{q1}B^{*}_{1i}+A^{*}_{q2}B^{*}_{2i}+..+A^{*}_{qn}B^{*}_{ni} \\
&=B^{*}_{1i}A^{*}_{q1}+B^{*}_{2i}A^{*}_{q2}+..+B^{*}_{ni}A^{*}_{qn}
\end{align*}
where we used equations (\ref{conj plus}) and (\ref{conj mult}) to
obtain the second line, and equation (\ref{commut}) to obtain the
third line. Now consider the components of
$B^{\dagger}A^{\dagger}$. By definition of matrix multiplication
we have that
\begin{align*}
(B^{\dagger}A^{\dagger})_{iq}&=B^{\dagger}_{i1}A^{\dagger}_{1q}+B^{\dagger}_{i2}A^{\dagger}_{2q}+..+B^{\dagger}_{in}
A^{\dagger}_{nq}\\
&=B^{*}_{1i}A^{*}_{q1}+B^{*}_{2i}A^{*}_{q2}+..+B^{*}_{ni}A^{*}_{qn}
\end{align*}
where the last line was obtained using the fact that
$A^{\dagger}_{ij}=A^*_{ji}$. Thus the components of
$(AB)^{\dagger}$ are precisely those of $B^{\dagger}A^{\dagger}$.
$\quad\square$

\begin{Pro}\label{probas} Let $V$ be a $n\times 1$ unit matrix of complex
numbers (a column). Then it is the case that:
\begin{align*}
|V_{11}|^2+|V_{21}|^2+..+|V_{n1}|^2=1
\end{align*}
\end{Pro}
\emph{Proof.} First let $z=(a,b)$ be a complex number, and note
that
\begin{align*}
z^{*}z&=(a^2+b^2,0) \\
&=(|z|^2,0)
\end{align*}
Now since $V$ is unit we have that
\begin{align*}
(V^{\dagger}V)_{11}&=V^{\dagger}_{11} V_{11}+V^{\dagger}_{12}
V_{21}+..+V^{\dagger}_{1n} V_{n1}\\
&=V^{*}_{11} V_{11}+V^{*}_{21} V_{21}+..+V^{*}_{n1} V_{n1}
\end{align*}
where we used successively: the definition of matrix
multiplication, and $A^{\dagger}_{ij}=A^*_{ji}$. The last line can
be further simplified using our first remark, namely:
\begin{align*}
V^{*}_{i1} V_{i1}=(|V_{i1}|^2,0)
\end{align*}
Thus
\begin{align*}
(V^{\dagger}V)_{11}&=(|V_{11}|^2,0)+(|V_{21}|^2,0)+..+(|V_{n1}|^2,0)\\
&=(|V_{11}|^2+|V_{21}|^2+..+|V_{n1}|^2,0)
\end{align*}
Because $V$ is unit the last line must be equal to $(1,0)$, and so
we have proved the property.$\quad\square$

\section{Quantum Theory}
Quantum theory is one of the pillars of modern physics. The theory
is $100$ years old and thoroughly checked by experiments; it
enables physicists to understand and predict the behaviors of any
closed (perfectly isolated from the rest of the world) physical
system. Usually these are small systems such as atoms, electrons,
photons etc. (only because they are generally less subject to
outside interactions).

\subsection{States}
\begin{Post}
\label{states}The \emph{state} of a \emph{closed physical system}
is wholly described by a unit $n\times 1$ matrix of complex
numbers.
\end{Post}

\emph{Comments.} In other words a state is given by a column of
$n$ complex numbers
\begin{align*}
V=\left(\begin{array}{c}
V_{11}    \\
\vdots\\
V_{n1}
\end{array}\right) \quad \textrm{such that}\quad V^{\dagger}V=I_1.
\end{align*}
What we mean by closed physical system is just about anything
which is totally isolated from the rest of the world. The number
of components $n$ varies depending on how complicated the system
is; it is called the \emph{degrees of freedom} or the
\emph{dimension} of the system. The postulate itself is extremely
short and simple. It is nonetheless puzzling as soon as you
attempt to
apprehend it with your classical intuition.\\
\emph{Example.} Consider a coin, which insofar as we have always
observed, can either by `head $\circledcirc$' or `tail
$\circledast$'. Thus we will suppose it has $n=2$ degrees of
freedom, and we will further assume that the state:
\begin{align*}
\textrm{`head}\;\circledcirc\textrm{' corresponds to quantum state
}\left(\begin{array}{c}
(1,0)  \\
(0,0)
\end{array}\right)\\
\textrm{whilst `tail}\;\circledast\textrm{'
 corresponds to quantum state }\left(\begin{array}{c}
(0,0)  \\
(1,0)
\end{array}\right)
\end{align*}
Now if the coin was to be shut in a totally closed box, it would
start behaving like a quantum coin. Thus the state:
\begin{align*}
\textrm{`}\circledcirc
+\circledast\textrm{'}=\left(\begin{array}{c}
(\frac{1}{\sqrt{2}},0)  \\
(\frac{1}{\sqrt{2}},0)
\end{array}\right)
\end{align*}
would become perfectly allowable. A quantum coin can be in a
\emph{superposition} of head and tail, i.e. it can be both head
and tail at the same time, in some proportion. Quantum theory is
more general than our classical intuition: it allows for more
possible states. It as if `head' and `tail' were two axes, and the
quantum coin was allowed to live in the plane described by those
axes.
\subsection{Evolution}
\begin{Post}
\label{evolution} A closed physical system in state $V$ will
evolve into a new state $W$, after a certain period of time,
according to
\begin{align*}
W=UV
\end{align*}
where $U$ is a $n\times n$ unit matrix of complex numbers.
\end{Post}

\emph{Comments.} In other words, in order to see how the quantum
state of a closed physical system evolves, you have to multiply it
by the matrix which describes its evolution (which we call $U$).
$U$ could be any matrix of complex numbers so long as it is
$n\times n$ (remember $V$ is an $n\times
1$ matrix) and verifies the condition $U^{\dagger}U=I_n$.\\
Note that this postulate is coherent with the first one, because
evolution under $U$ takes an allowed quantum state into an allowed
quantum state. Indeed suppose $V$ is a valid state, i.e. an
$n\times 1$ matrix verifying $V^{\dagger}V=I_1$. By definition of
the matrix multiplication an $n\times 1$ matrix multiplied by an
$n\times n$ matrix is also an $n\times 1$ matrix, and thus $W$ has
the right sizes. Is it a unit matrix? Yes:
\begin{align*}
W^{\dagger}W&=(UV)^{\dagger}(UV)\quad\textrm{by definition of W}\\
&=V^{\dagger}U^{\dagger}UV\quad\textrm{by Property \ref{dagger}}\\
&=V^{\dagger}I_n V\quad\textrm{since $U$ is unit}\\
&=V^{\dagger}V\quad\textrm{by Property \ref{identity}}\\
&=I_1\quad\textrm{since $V$ is unit}\\
\end{align*}
Thus $W$ is a valid quantum state.

\subsection{Measurement}
\begin{Post}
\label{measurement}When a physical system in state
\begin{align*}
V=\left(\begin{array}{c}
V_{11}  \\
\vdots\\
V_{n1}
\end{array}\right)
\end{align*}
is \emph{measured}, it yields outcome $i$ with probability
$p_i=|V_{i1}|^2$. Whenever outcome $i$ occurs, the system is left
in the state:
\begin{align*}
W=\left(\begin{array}{c}
(0,0)\\
\vdots\\
(1,0)\\
\vdots\\
(0,0)
\end{array}\right)\;\leftarrow\;i^{th}\textrm{ position}
\end{align*}
\end{Post}

\emph{Example.} Suppose you have a quantum coin in state:
\begin{align*}
\textrm{`}\circledcirc
+\circledast\textrm{'}=\left(\begin{array}{c}
(\frac{1}{\sqrt{2}},0)  \\
(\frac{1}{\sqrt{2}},0)
\end{array}\right)
\end{align*}
which you decide to measure. With a probability
$p_1=|\frac{1}{\sqrt{2}}|^2=\frac{1}{2}$ you will know that
outcome `1' has occurred, in which case your quantum system will
be left in state
\begin{align*}
\textrm{`}\circledcirc\textrm{'}=\left(\begin{array}{c}
(1,0)  \\
(0,0)
\end{array}\right)
\end{align*}
But with probability $p_2=\frac{1}{2}$ outcome `2' may occur
instead,in which case your quantum system will be left in state
`$\circledast$'.

\emph{Comments.} Thus a measurement in quantum theory is
fundamentally a probabilistic process. For this postulate to work
well we need to be sure that the probabilities all sum up to $1$
(so that something happens $100\%$ of the time). But you can check
that this is the case:
\begin{align*}
p_1+...+p_n&=|V_{11}|^2+..+|V_{n1}|^2 \quad\textrm{by postulate
\ref{measurement}}\\
&=1\quad\textrm{by Property \ref{probas}}
\end{align*}
The other striking feature of this postulate is that the state of
the system gets \emph{changed} under the measurement. In our
example everything happens as though the quantum coin in state
`$\circledcirc +\circledast$' is asked to make up its mind between
`$\circledcirc$' and `$\circledast$'. The quantum coin decides at
random, but once it does it remains coherent with its decision:
its
new state is either `$\circledcirc$' or `$\circledast$'.\\
This feature provides the basis for one of the latest high-tech
applications of quantum theory: quantum cryptography. Suppose
Alice and Bob want to communicate secretly over the phone, but
Eve, the Eavesdropper, might be spying upon their conversation.
What Alice and Bob can do is to send quantum coins to each other
across the (upgraded) phone network. As Eve attempts to measure
what the honest parties are saying, she is bound to \emph{change}
the state of the coin. This will enable\cite{BB84} Alice and Bob
to detect her malevolent presence.

\section{Deutsch-Jozsa algorithm}
The measurement postulate will (probably) make you think that
quantum theory is just a convoluted machinery whose only purpose
is to describe objects which might be in `state $1$' with
probability $p_1$, in `state $2$' with probability $p_2$ etc.
until $n$. After all why bother thinking of the state
`$\circledcirc +\circledast$' as a coin which is both head
`$\circledcirc$' and tail `$\circledast$' at the same time - when
after it gets observed it collapses to
either head `$\circledcirc$' or tail `$\circledast$' anyway? \\
No. You \emph{have} to consider that the coin is both
`$\circledcirc$' and `$\circledast$' \emph{until you measure it},
because this \emph{is} how it behaves \emph{experimentally} (until
you measure it). In other words the only way to account for what
happens between the moment you prepare your initial system and the
moment you measure it is to think of the complex components of the
state $V$ as \emph{amplitudes, proportions} and \emph{not} as
probabilities. This has much to do with what Postulate
\ref{evolution} enables us
to do.\\
In this last part we shall illustrate this point by considering
the simplest of all known quantum algorithms\cite{DJ}. An
\emph{algorithm} is just a recipe that is used to systematically
solve a mathematical problem. But the mathematical problem we will
now introduce cannot be solved by classical means: it can only be
solved using quantum theory, that is with a quantum algorithm. The
fact that this algorithm \emph{does work in practice} ought to
demonstrate the fact that the amplitudes of quantum theory permit
us to do things which mere probabilities would not allow, and
would not explain.

\subsection{The problem}
A \emph{boolean value} is something which can either be
$\mathbf{True}$ or $\mathbf{False}$. For instance the statement
`the sky is blue' has the boolean value $\mathbf{True}$ almost
anywhere in the world with the exception of
England, where it takes the value $\mathbf{False}$. \\
A \emph{boolean operator} is just a `box' which takes one or
several boolean values and returns one or several boolean values.
In order to define our problem we need to become familiar with two
boolean operators, which we now describe. \\
The boolean operator $\mathbf{Not}$ takes the boolean value
$\mathbf{True}$ into $\mathbf{False}$ and the boolean value
$\mathbf{False}$ into $\mathbf{True}$. We denote this as follows:
\begin{align*}
\mathbf{Not(True)}&=\mathbf{False} \\
\mathbf{Not(False)}&=\mathbf{True}
\end{align*}
The boolean operator $\mathbf{Xor}$ (exclusive or) takes two
boolean values and returns one boolean value. It returns
$\mathbf{True}$ either if the first boolean value it takes is
$\mathbf{True}$ and the second one is $\mathbf{False}$ or if the
second boolean value it takes is $\mathbf{True}$ and the first one
is $\mathbf{False}$. Otherwise it returns $\mathbf{False}$. We
denote this as follows:
\begin{align*}
\mathbf{Xor(True,False)}&=\mathbf{True} \\
\mathbf{Xor(False,True)}&=\mathbf{True} \\
\mathbf{Xor(False,False)}&=\mathbf{False} \\
\mathbf{Xor(True,True)}&=\mathbf{False}
\end{align*}
In other words $\mathbf{Xor}$ compares its two input boolean
values: it returns $\mathbf{True}$ if they are different and
$\mathbf{False}$ if they are the same.\\
We are now ready to state the problem.
\begin{Pb}
\label{xor pb} Suppose we are given a mysterious boolean operator
$\mathbf{F}$ (a black box) which takes one boolean value and
returns another boolean value. We want to calculate
$\mathbf{Xor(F(False),F(True))}$, i.e. the boolean value returned
by $\mathbf{Xor}$ when applied to the two possible results of
$\mathbf{F}$. But we are allowed to use the mysterious boolean
operator $\mathbf{F}$ only once.
\end{Pb}
It is clear that this problem cannot be solved classically. This
is because in order to learn anything about $\mathbf{F}$ you will
have to use $\mathbf{F}$. But we are allowed to do this only once.
Suppose we use $\mathbf{F}$ on input boolean value
$\mathbf{False}$. This gives us $\mathbf{F(False)}$, but tells us
nothing about $\mathbf{F(True)}$ which may still be either
$\mathbf{True}$ or $\mathbf{False}$. Thus we cannot compute
$\mathbf{Xor(F(False),F(True))}$ and we fail to solve the problem.
The same reasoning applies if we begin by using
$\mathbf{F}$ to obtain $\mathbf{F(True)}$.\\
But what would happen if we had the possibility to use
$\mathbf{F}$ upon an input boolean value which is both
$\mathbf{True}$ and $\mathbf{False}$, in some proportions (a
superposition)?

\subsection{The quantum setup}

Now suppose that the mysterious boolean operator $\mathbf{F}$ is
given in the form of a `quantum black box' instead. To make this
more precise we need to call
\begin{align*}
\textrm{`}\mathbf{False},\mathbf{False}\textrm{' the quantum
state}\left(\begin{array}{c}
(1,0)\\
(0,0)\\
(0,0)\\
(0,0)
\end{array}\right)\\
\textrm{`}\mathbf{False},\mathbf{True}\textrm{' the quantum
state}\left(\begin{array}{c}
(0,0)\\
(1,0)\\
(0,0)\\
(0,0)
\end{array}\right)\\
\textrm{`}\mathbf{True},\mathbf{False}\textrm{' the quantum
state}\left(\begin{array}{c}
(0,0)\\
(0,0)\\
(1,0)\\
(0,0)
\end{array}\right)\\
\textrm{`}\mathbf{True},\mathbf{True}\textrm{' the quantum
state}\left(\begin{array}{c}
(0,0)\\
(0,0)\\
(0,0)\\
(1,0)
\end{array}\right)
\end{align*}
We assume we have access, for one use only, to a physical device
which implements $\mathbf{F}$ as a quantum evolution. This quantum
evolution $U$ must take
\begin{align*}
\textrm{`}\mathbf{True,False}\textrm{' into }\textrm{`}\mathbf{True,F(True)}\textrm{'} \\
\textrm{`}\mathbf{False,False}\textrm{' into
}\textrm{`}\mathbf{False,F(False)}\textrm{'}
\end{align*}
Notice that if for instance `$\mathbf{F(True)}=\mathbf{True}$'then
`$\mathbf{True,F(True)}$' simply denotes the quantum state
`$\mathbf{True,True}$'. Furthermore we assume $U$ takes
\begin{align*}
\textrm{`}\mathbf{True,True}\textrm{' into }\textrm{`}\mathbf{True,Not(F(True))}\textrm{'} \\
\textrm{`}\mathbf{False,True}\textrm{' into }\textrm{`}\mathbf{False,Not(F(False))}\textrm{'} \\
\end{align*}
The quantum evolution $U$ is fully specified in this manner. In
matrix form it is given as follows:
\begin{align*}
\left(\begin{array}{cccc}
(1-F_\textrm{False},0)     &(F_\textrm{False},0)    &(0,0)     &(0,0)\\
(F_\textrm{False},0)     &(1-F_\textrm{False},0)   &(0,0)     &(0,0)\\
(0,0)     &(0,0)    &(1-F_\textrm{True},0)     &(F_\textrm{True},0)\\
(0,0)     &(0,0)   &(F_\textrm{True},0)     &(1-F_\textrm{True},0)
\end{array}\right)
\end{align*}
with:\\
$F_\textrm{False}$ equal to $1$ if $\mathbf{F(False)}$ is
$\mathbf{True}$, and $0$ otherwise.\\
$F_\textrm{True}$ equal to $1$ if $\mathbf{F(True)}$ is
$\mathbf{True}$
, and $0$ otherwise.\\
\\
Whatever the values of $F_\textrm{False}$ and $F_\textrm{True}$,
the matrix of complex number defined above is unit, i.e.
$U^{\dagger}U=I_4$. Thus according to postulate \ref{evolution}
this mysterious quantum black box is perfectly allowable
physically.\\
\emph{As an exercise you may want to check that the matrix $U$
does take `$\mathbf{True,False}$' into `$\mathbf{True,F(True)}$'
etc., and that it is indeed unit.}\\
\\
For our quantum algorithm we will need another quantum evolution:
\begin{align*}
H= \left(\begin{array}{cccc}
(1/2,0)     &(1/2,0)    &(1/2,0)     &(1/2,0)\\
(1/2,0)     &(-1/2,0)   &(1/2,0)     &(-1/2,0)\\
(1/2,0)     &(1/2,0)    &(-1/2,0)     &(-1/2,0)\\
(1/2,0)     &(-1/2,0)   &(-1/2,0)     &(1/2,0)
\end{array}\right)
\end{align*}
This $H$ is also a unit matrix of complex numbers.

\subsection{The solution}

\begin{Algo}
In order to solve problem \ref{xor pb} one may use the following
algorithm:\\
\\
$1$. Start with a closed physical system in quantum
state `$\mathbf{False,True}$'.\\
$2$. Evolve the system under the quantum evolution
$H$.\\
$3$. Evolve the system under the quantum evolution
$U$.\\
$4$. Evolve the system under the quantum evolution
$H$.\\
$5$. Measure the system.\\
\\
If $\mathbf{Xor(F(False),F(True))}$ is $\mathbf{False}$ the
quantum measurement always yields outcome `$2$'. \\
On the other hand if $\mathbf{Xor(F(False),F(True))}$ is
$\mathbf{True}$ the quantum measurement always yields outcome `$4$'.\\
Thus the algorithm always manages to determine
$\mathbf{Xor(F(False),F(True))}$, and does so with only one use of
the quantum evolution $U$.
\end{Algo}
\emph{Proof.} In Step $1$ we start with a closed physical system
whose quantum state is $V=\left(\begin{array}{c}
(0,0)\\
(1,0)\\
(0,0)\\
(0,0)
\end{array}\right)$.\\
After Step $2$ the quantum state of the system has become $HV$. By
working out this matrix multiplication we have
$HV=\left(\begin{array}{c}
(1/2,0)\\
(-1/2,0)\\
(1/2,0)\\
(-1/2,0)
\end{array}\right)$.\\
\emph{You may want to check this matrix multiplication and the
ones to follow, as an exercise.}\\
After Step $3$ the quantum state of the system has become $UHV$.
We can still work out the matrix multiplication but obviously the
result now depends upon our mysterious boolean operator
$\mathbf{F}$. Indeed we have $UHV=\left(\begin{array}{c}
(1/2-F_\textrm{False},0)\\
(-1/2+F_\textrm{False},0)\\
(1/2-F_\textrm{True},0)\\
(-1/2+F_\textrm{True},0)
\end{array}\right)$.\\
Notice that $UHV$ depends both upon $\mathbf{F(False)}$ and $\mathbf{F(True)}$, in some proportions. \\
After Step $4$ the quantum state of the system has become $HUHV$
and we have, by working out the multiplication:
$HUHV=\left(\begin{array}{c}
(0,0)\\
(1-F_\textrm{False}-F_\textrm{True},0)\\
(0,0)\\
(F_\textrm{True}-F_\textrm{False},0)
\end{array}\right)$.\\
Finally in Step $5$ we measure the state $HUHV$. According to
Postulate
\ref{measurement} this yields:\\
- outcome `$1$' with probability $0$ (never).\\
- outcome `$2$' with probability
$p_2=(1-(F_\textrm{False}+F_\textrm{True}))^2$.\\
- outcome `$3$' with probability $0$ (never).\\
- outcome `$4$' with probability
$p_4=(F_\textrm{True}-F_\textrm{False})^2$.\\
\\
Now if $\mathbf{Xor(F(False),F(True))}$ is $\mathbf{False}$ then
$F_\textrm{False}$ and $F_\textrm{True}$ have to be the same. Thus
$F_\textrm{False}+F_\textrm{True}$ equals either $0$ or $2$,
whereas $F_\textrm{True}-F_\textrm{False}$ is necessarily worth
$0$. As a consequence $p_2$ must equal $1$ whereas $p_4$ is worth
$0$.\\
Similarly, if $\mathbf{Xor(F(False),F(True))}$ is $\mathbf{True}$
then $F_\textrm{False}$ and $F_\textrm{True}$ have to be the
different values. Thus $F_\textrm{False}+F_\textrm{True}$ is
necessarily worth $1$, whereas $F_\textrm{True}-F_\textrm{False}$
equals either $-1$ or $1$. As a consequence $p_2$ is worth $0$
whereas $p_4$ must equal $1$. $\quad\square$

\subsection{Comments}
It is quite a remarkable fact that with only one use of the
`quantum black box' we succeed to determine a quantity which
intrinsically depends `on both possible values which the box may
return'. Although this algorithm does not seem extremely useful in
every day life, it teaches us an important lesson: the components
of a quantum state must be viewed as proportions (amplitudes), not
as probabilities. The quantum coin can be both head or tail in
some proportions, simultaneously, until you measure it.\\
Until recently this feature of quantum theory was essentially
regarded as an unfortunate oddity which made the theory difficult
to grasp. But we are now learning to turn this feature to our own
advantage, as a means of `exploring several possibilities
simultaneously' (so to speak).\\
This is recent research however, and to this day not so many
quantum algorithms are known. Yet we do know that Quantum
Computers can factorize large integer numbers efficiently, or even
find a name within an unordered list of $100$ people in only $5$
tries. These are quite useful things to be able to do. The best
place to learn about them is \cite{Nielsen}, if you have followed
me this far you can go further.

\section{Acknowlegments}
The author would like to thank his mother for suggesting this
article, Anuj Dawar for his patient listening, EPSRC, Marconi, the
Cambridge European and Isaac Newton Trusts for financial support.

\end{document}